\documentclass[nofootinbib,showpacs,showkeys,prd,superscriptaddress]{revtex4}

\usepackage{amsmath}
\usepackage{amsfonts}
\usepackage{amssymb}
\usepackage{graphicx}

\begin{document}


\title{Curvature dark energy reconstruction through different cosmographic distance definitions}

\author{Salvatore Capozziello}
\email{capozzie@na.infn.it}
\affiliation{Dipartimento di Fisica, Universit\`a di Napoli ''Federico II'', Via Cinthia, I-80126 Napoli, Italy.}
\affiliation{Istituto Nazionale di Fisica Nucleare (INFN), Sez. di Napoli, Via Cinthia, I-80126 Napoli, Italy.}
\affiliation{Gran Sasso Science Institute (INFN), Viale F.  Crispi 7,   I-67100  L'Aquila, Italy.}

\author{Mariafelicia De Laurentis}
\email{felicia@na.infn.it}
\affiliation{Dipartimento di Fisica, Universit\`a di Napoli ''Federico II'', Via Cinthia, I-80126 Napoli, Italy.}
\affiliation{Istituto Nazionale di Fisica Nucleare (INFN), Sez. di Napoli, Via Cinthia, I-80126 Napoli, Italy.}
\affiliation{Tomsk State Pedagogical University, 634061 Tomsk and National Research Tomsk State University, 634050 Tomsk, Russia.}

\author{Orlando Luongo}
\email{luongo@na.infn.it}
\affiliation{Dipartimento di Fisica, Universit\`a di Napoli ''Federico II'', Via Cinthia, I-80126 Napoli, Italy.}
\affiliation{Istituto Nazionale di Fisica Nucleare (INFN), Sez. di Napoli, Via Cinthia, I-80126 Napoli, Italy.}
\affiliation{Instituto de Ciencias Nucleares, Universidad Nacional Autonoma de M\'exico (UNAM), Mexico.}

\begin{abstract}
In the context of $f(\mathcal{R})$ gravity, dark energy is   a geometrical fluid with negative equation of state. Since the function $f(\mathcal{R})$ is not known \emph{a priori}, the need of a model independent reconstruction of its shape   represents a relevant technique to determine which $f(\mathcal{R})$ model  is really favored with respect to others. To this aim, we relate cosmography to a generic $f(\mathcal R)$ and its derivatives in order to provide a model independent investigation at redshift $z \sim 0$. Our analysis is  based on the use of three different cosmological distance definitions, in order to alleviate the duality problem, i.e. the problem of which cosmological distance to use with specific cosmic data sets. We therefore consider the luminosity, $d_L$, flux, $d_F$, and angular, $d_A$, distances and we find numerical constraints by the Union 2.1 supernovae compilation and measurement of baryonic acoustic oscillations, at $z_{BAO}=0.35$.  We notice that all distances reduce to the same expression, i.e. $d_{L;F;A}\sim\frac{1}{\mathcal H_0}z$, at first order. Thus, to fix the cosmographic series of observables, we impose the initial value of  $H_0$ by fitting $\mathcal H_0$  through supernovae only, in the redshift regime $z<0.4$. We find that the pressure of curvature dark energy fluid is slightly lower than the one related to the cosmological constant. This indicates that a possible evolving curvature dark energy realistically fills the current universe. Moreover, the combined use of $d_L,d_F$ and $d_A$ shows that the sign of the acceleration parameter agrees with theoretical bounds, while its variation, namely the jerk parameter, is compatible with $j_0>1$. Finally, we infer the functional form of $f(\mathcal{R})$ by means of a truncated polynomial approximation, in terms of fourth order scale factor  $a(t)$.
\end{abstract}


\maketitle

\section{Introduction}
\label{sec:int}

Modern cosmology is nowadays plagued by several shortcomings which jeopardize the current understanding of universe dynamics. Particularly, these problems may suggest  to reconsider the standard approach of gravitation, based on Einstein's gravity, in favor of alternative theories of gravity. Alternative gravity pictures have been extensively introduced in order to describe universe dynamics without the need of additional material ingredient as dark energy and  dark matter described by new particles at fundamental level.  On the other hand, the simple introduction of a cosmological constant vacuum energy  seems  inadequate to characterize the whole universe evolution at any epoch \cite{adj}. Thus, these alternative theories are viewed as a bid to reformulate \emph{in toto} semi-classical schemes where  General Relativity is only a particular case of a more extended theory.  In particular, such theories  are  able to extend General Relativity predictions by means of higher order curvature invariants. Other pictures assume extensions based on minimally or non-minimally coupled scalar fields in the gravitational Lagrangians \cite{review1,review2,review3,review4,review5}. Furthermore,  Einstein gravity can be extended  by carrying out the full Mach principle: this fact leads to the introduction  of a varying gravitational coupling. Under these hypotheses, the Brans-Dicke theory \cite{BD}  represents the prototype of  alternative schemes to General Relativity. It naturally includes  a variable gravitational coupling, whose dynamics is governed by a single scalar field non-minimally coupled to geometry \cite{BD, cimento,odintsov}. From  one hand, extensions of General Relativity are therefore able to  describe the above-mentioned theoretical aspects. On the other hand, it is also possible to account every unification scheme of fundamental interactions, such as superstring, supergravity, or grand unified theories and physically by low-energy effective actions containing non-minimal couplings  or higher order curvature terms \cite{sciama}. In fact, interactions between quantum scalar fields and background geometry, or gravitational self-interactions, naturally yield such corrections to the Einstein-Hilbert's Lagrangian \cite{birrell}. Hence, it is easy to show that several geometrical corrections  are inescapable within quantum gravity effective actions and allow consistent  pictures  close to Planck  scales \cite{vilkovisky}. These schemes represent working approaches towards a self consistent quantum picture, giving rise to interesting consequences once corrections like $\mathcal{R}^{2}$, $\mathcal{R}^{\mu\nu}$, $\mathcal{R}_{\mu\nu}$, $\mathcal{R}^{\mu\nu\alpha\beta}\mathcal{R}_{\mu\nu\alpha\beta}$, $\mathcal{R} \,\Box \mathcal{R}$, or $\mathcal{R}\,\Box^{k}\mathcal{R}$ are involved. A crucial fact is that alternative theories may provide analogies with the effective string or Kaluza-Klein Lagrangians, when compactification mechanisms  of extra spatial dimensions are imposed \cite{veneziano}.

A consequence of such extended theories of gravity is the possibility to frame current universe dynamics in a self consistent way considering their infrared counterpart. In particular, such models may address the problem of  current universe speed up \cite{sn} by considering further gravitational  degrees of freedom. Indeed, General Relativity seems not capable  of dealing  with  present cosmic acceleration, unless an unknown fluid dubbed dark energy is added  to the standard matter fluid energy-momentum tensor. At late times, the fluid responsible for accelerating the universe  dominates over all other contributions, driving the universe evolution.  It should be able to reproduce current observations \cite{gala}. Consequently, the dark energy equation of state behaves anti-gravitationally by counterbalancing gravitational attraction \cite{ioequew}. Thus,  in this {\it concordance model},  the universe dynamics is  described through pressureless matter terms, i.e. the sum of baryons and  cold dark matter,    through a  evolving barotropic   dark energy contribution and a vanishing spatial curvature $\Omega_k=0$ \cite{qualcosa1,qualcosa2, qualcosa3}.

A straightforward way to address geometrically  the problem of dark energy is by the so called $f(\mathcal{R})$ gravity, where $f$ is a generic function of the Ricci scalar ${\mathcal R }$ \cite{review3,review4,review5}.
In this paper, we fix constraints on geometrical dark energy fluid inferred in the context of $f(\mathcal{R})$ gravity. To this end, we adopt  cosmography to fix cosmological bounds on the $f(\mathcal{R})$ function and its derivatives at low redshift regime where degeneracy of concurring dark energy models is more evident. Cosmography allows to determine cosmological constraints in a model-independent way, once scalar curvature is somehow fixed. The idea is to expand into Taylor series cosmological observables. These expansions can be compared with data to get the cosmographic series, i.e. the numerical bounds on scale factor derivatives \cite{mio,salzano1,salzano2}. One commonly-used technique is represented by expanding the luminosity distance and compare it with supernovae data. However, a degeneration  problem (duality problem)  occurs once different cosmological distances are involved. Hence, a non-definitive consensus exists on the  adequate cosmological distance to use in the framework of cosmography. We therefore perform the experimental analysis by means of three cosmological distance rulers, i.e. luminosity, flux and angular distances. We check the viability of different cosmological distances and  measure cosmological constraints on the cosmographic series, deriving bounds on $f(\mathcal{R})$ curvature dark energy.

The paper is organized as follows: in Sec. II, we highlight the main features of cosmography and its application to cosmology. In Sec. III, we describe the problems related to cosmography, pointing out  the so called duality problem. In Sec. IV,  the experimental procedures is described. In Sec. V, cosmography in view of  $f(\mathcal{R})$ is discussed . Finally, Sec. VI is devoted to conclusions and perspectives.

\section{Basics of Cosmography}

Let us summarize the main aspects of cosmography and  describe how it can be considered as a tool to fix constraints on cosmological observables. Firstly, let us  assume that the cosmological principle holds and the  equation of state is currently determined by a geometrical fluid, with  pressure $P_{curv}$. Under these hypotheses, we expand cosmological observables into Taylor series and match the derivatives of such expansions with cosmological data. Examples of expanded quantities are  the Hubble parameter, the luminosity distance, the apparent magnitude modulus  \cite{orl,orl2}, the net pressure, and so forth \cite{orl21,orl99}. The power series coefficients of the scale factor expansion are known in the literature as \emph{cosmographic series} (CS), if calculated at present time, or alternatively at the redshift $z=0$. Those coefficients are therefore expressed in terms of the cosmological scale factor $a(t)$ and its derivatives \cite{mio}. It follows that the cosmographic approach does not need to assume a particular cosmological model.

Thence, one of the main advantage of cosmography is alleviating degeneracy among cosmological models, i.e. cosmography allows, in principle,  to understand which model better behaves than others. In case of $f(\mathcal{R})$ gravity, for example, matter density degenerates with scalar curvature and cannot be constrained \emph{a priori}. However, cosmography fixes model independent constraints on the cosmological  equation of state and then results  a technique to discriminate among competing $f(\mathcal{R})$ models \cite{mio}, removing the  degeneracy between matter and scalar curvature \cite{salzano1}. This technique turns out to be useful to reconstruct the form of $f(\mathcal R)$ which better traces the universe expansion history. Thus, more precisely, cosmography represents a  model independent method to infer cosmological bounds, once spatial curvature is somehow fixed.

Recent observations point out that the scalar curvature is negligible,  so we can easily impose $\Omega_k=0$ \cite{ob}. Than one has
\begin{eqnarray}\label{serie1a}
\frac{1-a(t)}{H_0} & \sim &   \Delta t + \frac{q_0}{2} H_0 \Delta t^2 +
\frac{j_0}{6} H_0^2 \Delta t^3 -   \frac{s_0}{24} H_0^3 \Delta t^4 +\ldots\,,
\end{eqnarray}
which represents the Taylor series of the scale factor $a(t)$, around $\Delta t= t-t_0=0$. The CS can be thus defined as
\begin{eqnarray}\label{eq:CSoftime}
\frac{\dot{H}_0}{H_0^2}=-(1+q)\,, \quad \frac{\ddot{H}_0}{H_0^3}=j+3q+2\,, \quad \frac{H_0^{(3)}}{H_0^4}=s-4j-3q\left(q+4\right)-6\,.
\end{eqnarray}
Here, dots represent derivatives with respect to the cosmic time $t$. Each term brings  its own  physical meaning. Particularly, the Hubble rate $H(t)$ is intimately related to the variation of $a(t)$ with time, the acceleration parameter $q(t)$ measures how the universe is speeding up and the jerk parameter $j(t)$ permits one to understand how the acceleration varied in the past. The coefficients are defined as
\begin{equation} \label{eq:CScoeff}
     H(t) = \frac{1}{a}\frac{da}{dt}, \quad
    q(t) = -\frac{1}{a     H^2} \frac{d^2a}{dt^2}, \quad
    j(t) = \frac{1}{a     H^3} \frac{d^3a}{dt^3}\,,
\end{equation}
and are considered at a given time $t_0$.
We may argue that such quantities are able to describe the kinematics of the universe \cite{turnercosmografia} and we do not pose, at this stage,  the problem of which model causes the universe acceleration. In analogy to the classical mechanics, we say that cosmography is a kinematic approach to trace the universe expansion today.
From one hand, the advantages of cosmography consist  on its  model independent reconstructions of present-time cosmology. In other words, it can be considered like a snapshot of the today observed universe capable of giving initial conditions for reconstructing back the cosmic evolution.   From the other hand, the disadvantages rely on the fact that current data are either not enough to guarantee significative and accurate constraints or do not fit significant intervals of convergence for  $z\ll1$. In addition, the  cosmological observable that one  expands into Taylor series, i.e. $a(t)$  is not known \emph{a priori}. Consequently, there is  no  physical motivations to use a particular cosmological distance than others. This means that the use of a given  luminosity distance to constrain CS is only motivated  by \emph{ad hoc} arguments. This fact constitutes the so-called  \emph{duality problem} that we discuss in the next section. To alleviate duality problem, we will compare three different cosmological distances to trace universe expansion history at late times, under the hypothesis of a  $f(\mathcal{R})$ geometrical dark energy fluid.

\section{The duality problem and cosmographic convergence}

By a cosmographic analysis, one can  fix constraints on the geometrical dark fluid, alleviating the degeneracy problem. To this end,  one needs  a self-consistent  definition of causal distance. Unfortunately, standard  definitions  implicitly postulate that the universe is accelerating \cite{lix1,lix2}, i.e. to infer the distance expansion, we evaluate the distance $r_0$ that a photon travels from a light source at $r=r_0$ to our position at $r=0$, defined as ${\displaystyle r_0 = \int_{t}^{t_0}{\frac{dt'}{a(t')}}}$. Consequently, one obtains as prototype the so called luminosity distance $d_L$, while other definitions, e.g. the photon flux distance $d_F$, angular diameter distance $d_A$ and so forth, can easily be derived from different considerations. As previously stressed, this leads to a severe duality problem on the choice of the particular cosmological ruler to use for fixing cosmological constraints on the CS.

Here, we use  three different cosmological distances as rulers, e.g. the luminosity, flux and angular distances, $d_L$, $d_F$ and $d_A$ respectively. Below the definition of these distances is reported in terms of $r_0$, that is
\begin{subequations}\label{gb}
\begin{align}
    d_L  &=  a_0 r_0 (1+z) = r_0\,a(t)^{-1}\,,  \\
    d_F  &=  \frac{d_L}{(1+z)^{1/2}} = r_0\,a(t)^{-\frac{1}{2}}\,, \\
    d_A  &=  \frac{d_L}{(1+z)^2} = r_0\, a(t)\,.
\end{align}
\end{subequations}
These distances can be used to the  fix causal constraints on the curvature fluid  in order to alleviate the degeneracy problem. For the sake of clearness, it is important  to stress that although $d_L$ is associated to the ratio of the apparent and absolute luminosity of astrophysical objects, the other distances, i.e. $d_F$ and $d_A$, may be also used to fix bounds on the observable universe. All the different cosmological distances rely on the fundamental assumption that the total number of photons is conserved at cosmic scales. Hence, there is no reason to discard one distance with respect to another  since all of them fulfill this condition. The duality problem represents a  not well understood issue of observational cosmology \cite{bll}. In this work, we find  differences in  fitting Eqs. ($\ref{gb}$), showing that there is no reason to adopt $d_L$ only as the only cosmological  distance.

 However a problem of \emph{convergence} may occur, leading to possible misleading results for $z>1$ in the cosmographic Taylor series. An immediate example is due to the most high supernova redshift in a typical data set. Usually, one has that the furthest  redshift at approximatively $z\sim 1.41$, showing that a few number of supernovas spans in the range $z>1$. It follows that numerical divergences and bad convergences may occur in the analysis, since  Taylor expansions are carried out around $z=0$. A plausible landscape deals with introducing alternative redshift definitions, re-parameterizing the cosmological distances in a tighter redshift range \cite{gtm}. These possible re-parameterizations must fulfill the conditions that the distance curves should not behave too steeply in the interval $z<1$. Moreover, the luminosity distance curve should not exhibit sudden flexes, being one-to-one invertible as discussed in \cite{mio}. In other words, it is easy to show that the new redshift re-parameterization, i.e. $z_{new}$, provides  $z_{new}=\mathcal{Z}(z)$, with $\mathcal{Z}$ a generic function of the redshift $z$, with the property $\mathcal{Z}\rightarrow 1$, as $z\rightarrow\infty$. In this work, we describe a technique to reduce the convergence problem, calibrating cosmological distance at first order in the Taylor series within a smaller range of redshift. Our strategy is to fix $H_0$ with supernovae in the range $z<0.4$. This turns out to be useful since a wide range of data is actually inside the sphere $z<1$ and all cosmological distances at first order reduce to

\begin{equation}\label{diconniente}
d_i\sim\frac{z}{H_0}\,,
\end{equation}
where $d_i$ represents the generic distance, i.e. $i=L;F;A$. Once $H_0$ is fixed, the series naturally converges better  since its shape increases or decreases as $H_0$ decreases or increases respectively. In other words, the dynamical shape of any cosmological curve depends on the value given to $H_0$. As $H_0$ is somehow known,  curves behave better  at higher redshift, alleviating convergence problems as expected. These arguments represent a further tool  in order to fix model independent constraints on $f(\mathcal{R}(z))$ and its derivatives. Indeed, $H_0$ is fixed regardless the cosmological distance taken into account, by means of Eq. ($\ref{diconniente}$).
It is possible to  fix $H_0$ in the range $z<0.4$ with supernovae only. We find
\begin{equation}\label{gdixconi}
    H_0=69.785^{+1.060}_{-1.040}\,.
\end{equation}
In cosmography, the strategy of fixing $H_0$ in a smaller interval of data overcomes several problems associated to the well consolidated usage of auxiliary variables. Indeed, as above mentioned, the method of adopting auxiliary variables consists in determining parametric functions $y(z)$ in terms of the redshift $z$, whose values rely in the interval $y(z)\in[0,1]$, as $z\rightarrow0$ and $z\rightarrow \infty$ respectively. This procedure rearranges catalog data and suffers from severe shortcomings \cite{gtm,gtm2}. Indeed, the form of $y(z)$ is not known \emph{a priori} and any possible reparameterized variable should guarantee that errors do not deeply propagate in the statistical analysis. In several cases, $y(z)$ variables are therefore inconsistent with low redshift cosmography, providing misleading results, albeit their use becomes more relevant for high redshift data sets.

In our case, we propose to fix $H_0$ as a \emph{low redshift cosmographic setting value}, since all distances reduce to Eq. (\ref{diconniente}) at a first order of Taylor expansions. Our corresponding best fit intervals are compatible with previous analysis \cite{nfn21} and guarantee that errors do not significatively propagate on measured coefficients. For our purposes, the strategy of fixing $H_0$ by means of small redshift data only better behaves than standard auxiliary variables, due to the fact that $z\leq1$ (see for recent applications \cite{nfn22,ratra,ratra1}), although it would fail at higher redshift domains. Since, in our cases, $z$ reaches the upper value $z\sim 1.414$, i.e. the maximum $z$ of the supernova compilation, we expect that the use of Eq. (\ref{diconniente}) to get $H_0$ would guarantee refined best fit results with respect to any possible reparameterized auxiliary variables.

An additional technique is to combine more than one data set to infer cosmological bounds. Indeed, although we treat bad convergence of truncated series by using Eq. ($\ref{gdixconi}$), further data sets would improve the quality of numerical estimates. Hence, the CS and the derived constraints on $f(\mathcal R)$ derivatives would improve consequently. Here, we combine supernovae data with the baryonic acoustic oscillation measurement. It is possible to show  that such a  choice actually reduces the convergence problem.

\section{Cosmological data sets and the fitting procedure}

In this section, we describe the two data sets used for the cosmographic analysis. First, let us consider  the Union 2.1 compilation \cite{kow}. Second, we assume the measurement of baryonic acoustic oscillation (BAO) \cite{sn32}. As it is well known, supernovae data span in the plane $\mu-z$, consisting of 580 supernovae, in the observable range $ 0.015 < z < 1.414$. For our purposes, to fix viable constraints, we follow a standard Bayesian analysis, dealing with the determination of best fits, evaluated by maximizing the  likelihood function ${\mathcal L} \propto \exp(-\chi^2/2)$. Here, $\chi^2$ is  the  ({\it pseudo}){\it $\chi$-squared} function, or reduced $\chi$ squared. The distance modulus $\mu$ for each supernova is
\begin{equation}
\mu = 25 + 5 \log_{10} \frac{d_L}{Mpc}\,,
\end{equation}
and once given the corresponding  $\sigma_i$ error, we are able to minimize the $\chi$ square as follows
\begin{equation}
\chi^{2}_{SN} =
\sum_{i}\frac{(\mu_{i}^{\mathrm{theor}}-\mu_{i}^{\mathrm{obs}})^{2}}
{\sigma_{i}^{2}}\,.
\end{equation}
On the other hand, the large scale galaxy clustering observations provide the signatures for the baryonic acoustic oscillation.  This gives a further tool to explore the parameter space and alleviate the convergence problem. We use the peak measurement of luminous red galaxies observed in Sloan Digital Sky Survey (SDSS). By employing $\mathcal{A}$ as the measured quantity, we have
\begin{equation}
\mathcal{A}=\sqrt{\Omega_m}  \Big[\frac{H_0}{H(z_{BAO})}\Big]^{\frac{1}{3}}
\left[ \frac{1}{z_{BAO}}\int_0^{z_{BAO}}
\frac{H_0}{H(z)}dz\right]^{\frac{2}{3}}\,,
\end{equation}
with $z_{BAO}=0.35$. In addition, the observed $\mathcal{A}$ is estimated to be $\mathcal{A}_{obs} = 0.469 \left(\frac{0.95}{0.98}\right)^{-0.35}$, with an error $\sigma_\mathcal{A} = 0.017$. In the case of the BAO measurement, we minimize the $\chi$ squared
\begin{equation}
\chi^{2}_{BAO}=\frac{1}{\nu}\left(\frac{\mathcal{A}-\mathcal{A}_{obs}}{\sigma_\mathcal{A}}\right)^2\,.
\end{equation}
An important feature of BAO is that it does not depend on $H_0$.

 Estimations of the cosmographic parameters may be performed passing through the standard Bayesian technique, maximizing the likelihood function:
\begin{equation}\label{LIKER}
    \mathcal{L}_i
\propto \exp (-\chi_i^2/2 )\,,
\end{equation}
where $\chi_i^2$ is explicitly determined for
each compilations here employed and the subscript indicates the data set, i.e. supernovae or BAO. Maximizing the likelihood function is equivalent to minimizing the total $\chi_t\equiv\chi_{SN}+\chi_{BAO}$-squared
function and so one argues to maximize 
\begin{equation}\label{LIKER2}
    \mathcal{L}_{tot}\equiv\mathcal{L}_{SN}\times\mathcal{L}_{BAO}
\propto \exp (-\chi_t^2/2 )\,.
\end{equation}
In particular, the cosmographic results have been obtained by directly employing Eq. (\ref{LIKER2}) maximizing the corresponding likelihood functions over a grid, through a standard Bayesian analysis. In so doing, the cosmographic series has been evaluated and the provided errors refer to as the $1\,\sigma$, associated to the $68\%$ of confidence level. Once all cosmographic coefficients have been determined through a direct Gaussian maximization of the likelihood function, we will infer the derived coefficients, $f_0,f_{z0},f_{zz0},P_{curv}$, by simply propagating the errors through the well consolidated logarithmic method.

\section{$f(\mathcal{R})$ cosmography vs redshift}

\begin{figure*}
\begin{center}
\includegraphics[width=2in]{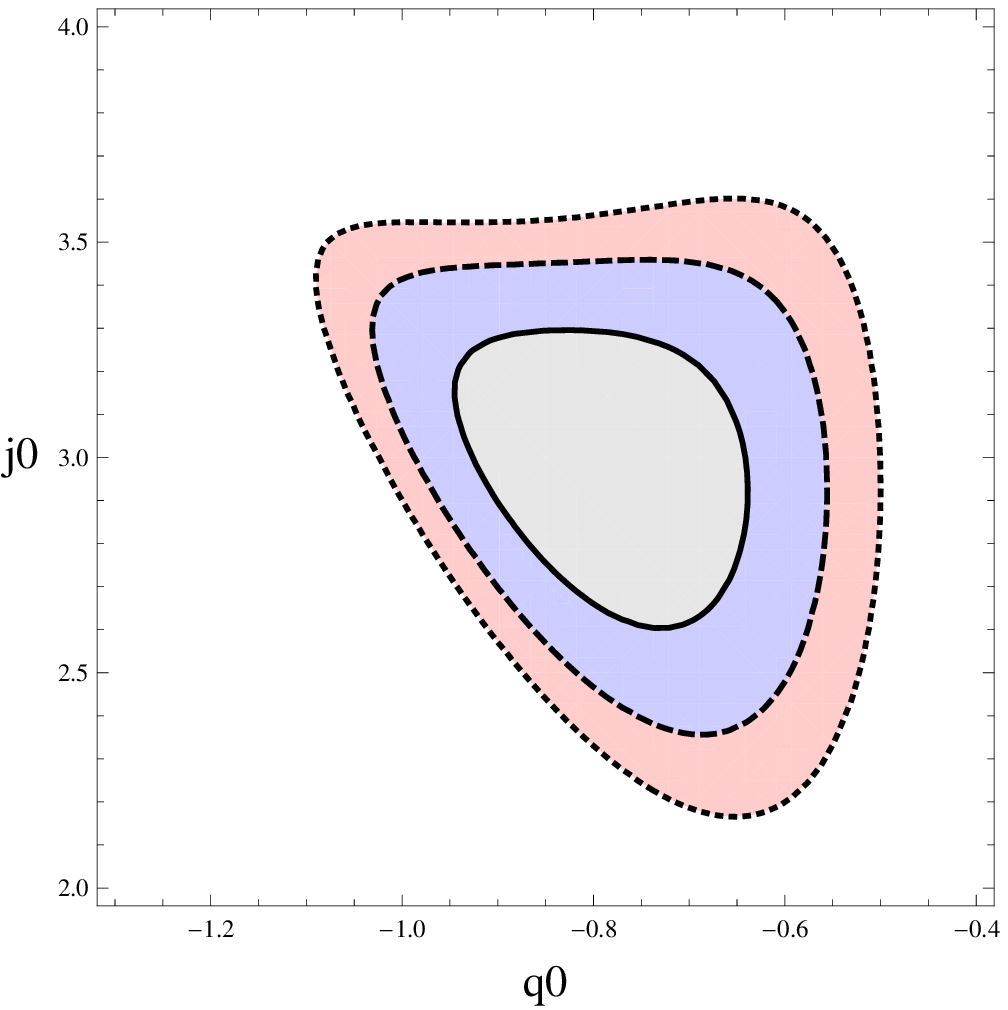}
\includegraphics[width=2in]{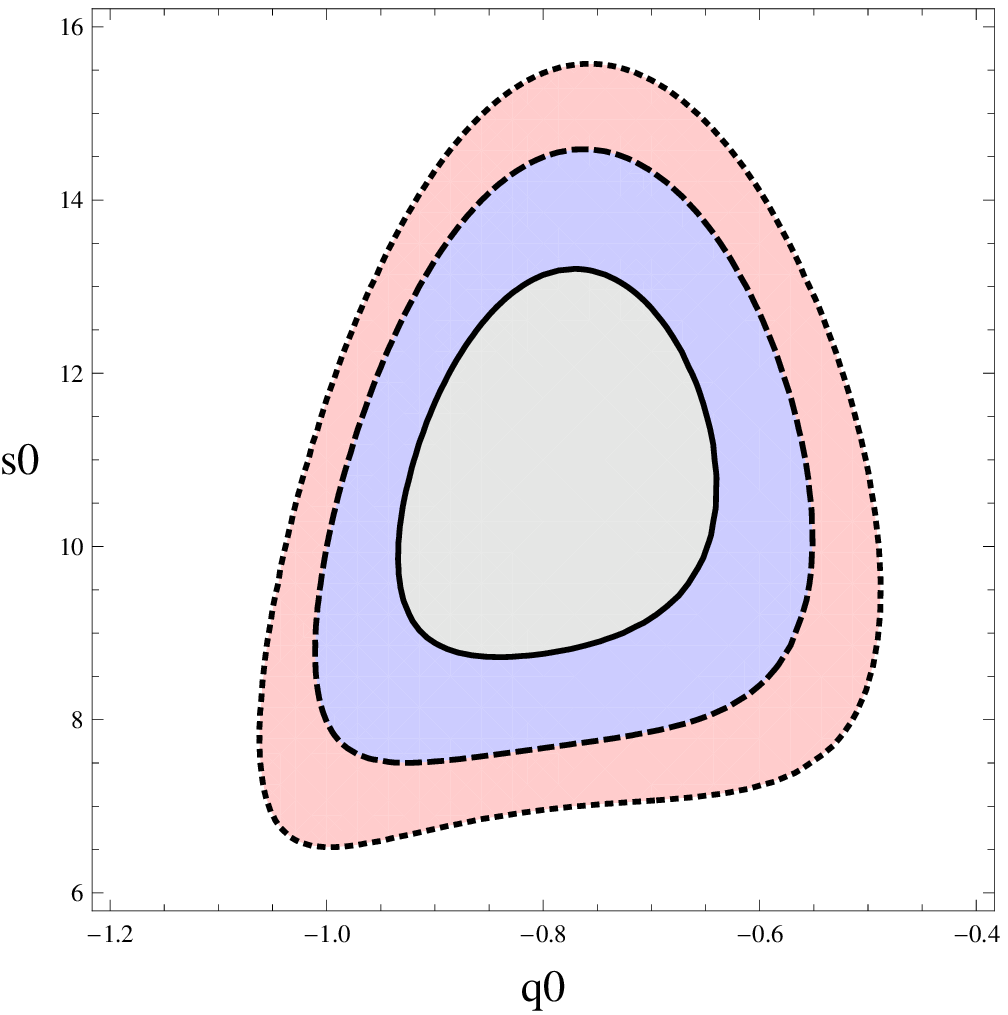}
\includegraphics[width=2in]{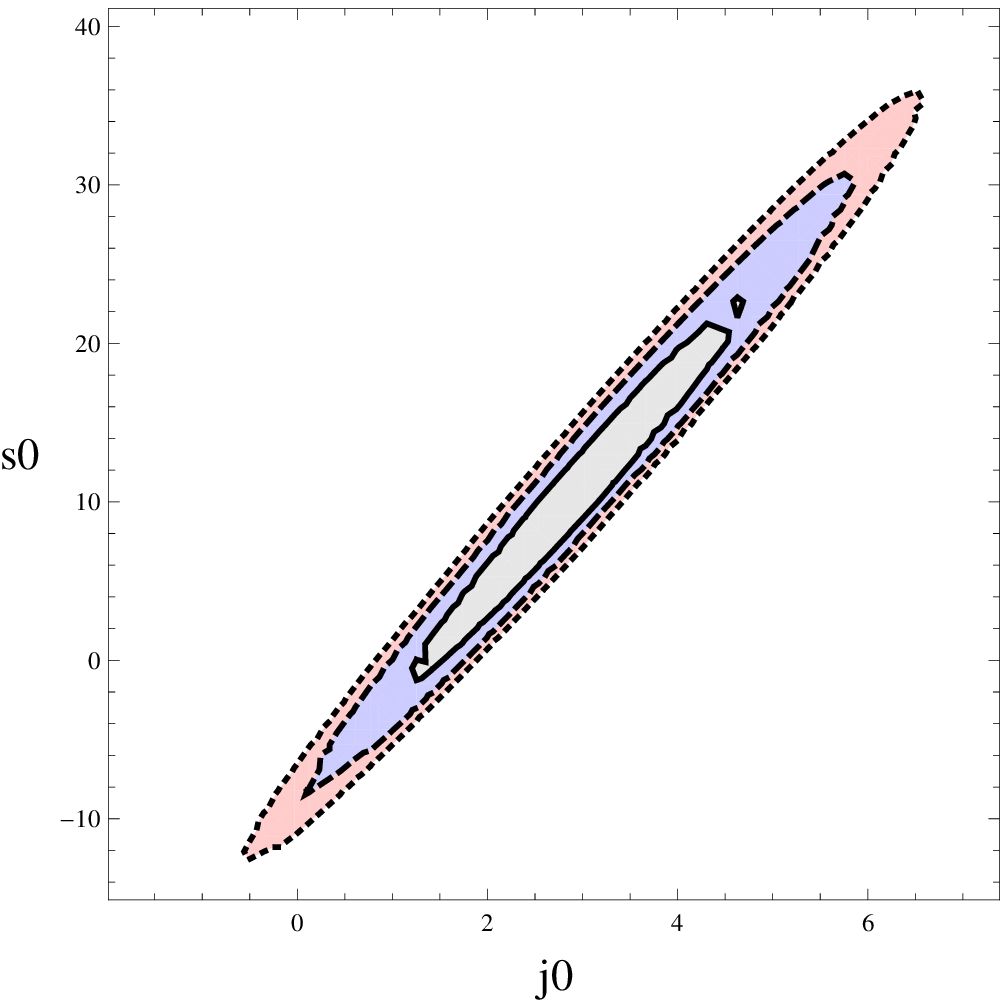}
\includegraphics[width=2in]{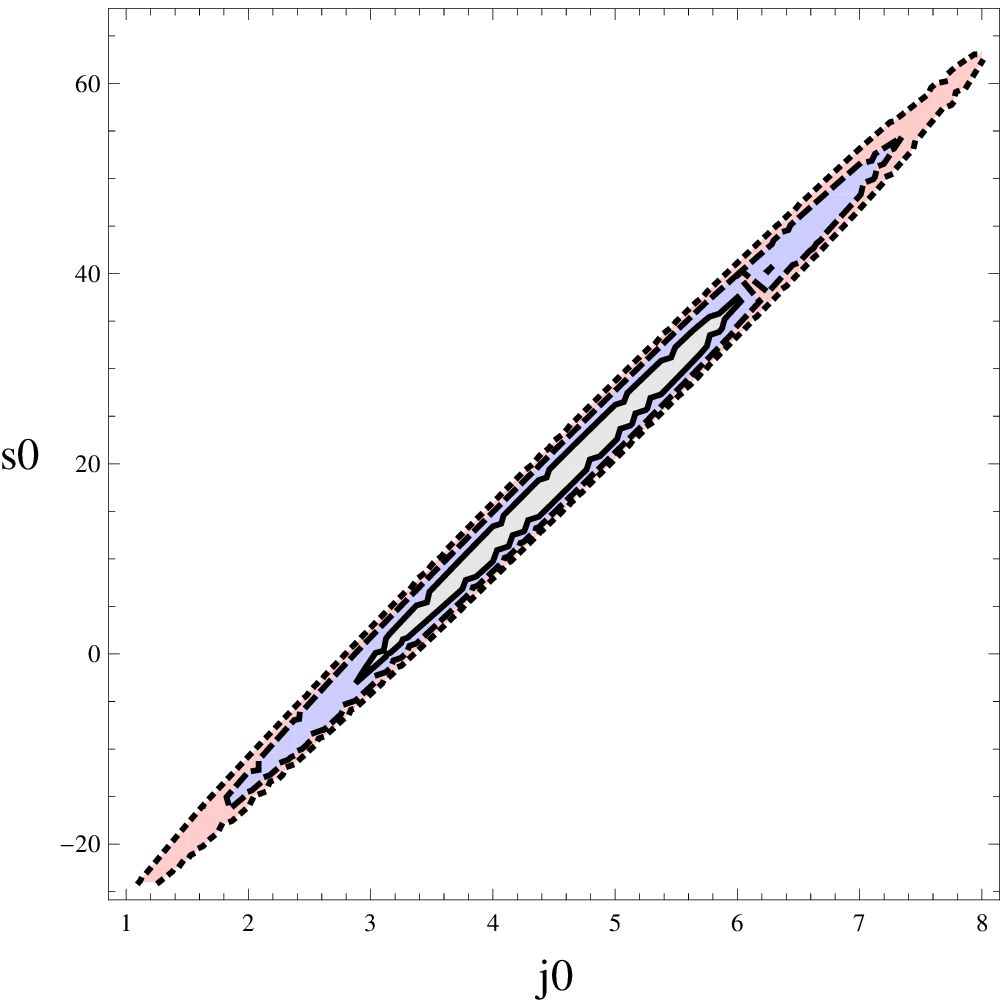}
{\small \caption{Contour plous derived from  fitting procedures. Above, as examples, we report the first three contour plots, obtained by using the luminosity distance and the last plot, derived from the use of $d_F$. Each plot shows the $68\%$, $95\%$ and $99\%$ of confidence levels. Easily one can notice that there are some difficulties in bounding $q_0$ versus $j_0$, as reported in Tables I and II, with $1\sigma$ error bars.}
\label{Fig:EoS}}
\end{center}
\end{figure*}

Let us consider now the $f(\mathcal{R})$ coefficients (evaluated in terms of the redshift $z$) as function of observable quantities. The robustness of  calculations leads to the  advantage of relating derivatives of $f(\mathcal{R}(z))$  at $z=0$ to experimental bounds, without assuming a priori a form of $f(\mathcal{R}(z))$. Hence, by expanding the causal distances $d_L$, $d_F$ and $d_A$, through the definition of the scale factor in terms of redshift, $a\equiv(1+z)^{-1}$, we get
\begin{eqnarray*}\label{dlinterminidiz}
d_L &=&  \frac{1}{H_0} \cdot \Bigl[ z + z^2 \cdot \Bigl(\frac{1}{2} - \frac{q_0}{2} \Bigr) +
    z^3 \cdot \Bigl(-\frac{1}{6} -\frac{j_0}{6} + \frac{q_0}{6} + \frac{q_0^2}{2} \Bigr)+ \nonumber\\
    &&+\, z^4 \cdot \Bigl( \frac{1}{12} + \frac{5 j_0}{24} - \frac{q_0}{12} + \frac{5 j_0 q_0}{12} -
    \frac{5 q_0^2}{8} - \frac{5 q_0^3}{8} + \frac{s_0}{24} \Bigr)+\ldots\Bigr]\,,
\end{eqnarray*}

\begin{eqnarray*}
    d_F & = & \frac{1}{H_0} \cdot \Bigl[ z - z^2 \cdot \frac{q_0}{2} + z^3 \cdot \Bigl(-\frac{1}{24}
    -\frac{j_0}{6} + \frac{5q_0}{12} + \frac{q_0^2}{2} \Bigr)+ \nonumber\\
    &&+\, z^4 \cdot \Bigl( \frac{1}{24} + \frac{7 j_0}{24} - \frac{17 q_0}{48} + \frac{5 j_0 q_0}{12}
    - \frac{7 q_0^2}{8} - \frac{5 q_0^3}{8} + \frac{s_0}{24} \Bigr)+\ldots\Bigr]\,,
\end{eqnarray*}
and
\begin{eqnarray*}
    d_A & = & \frac{1}{H_0} \cdot \Bigl[ z + z^2 \cdot \Bigl( -\frac{3}{2} - \frac{q_0}{2} \Bigr)
    + z^3 \cdot \Bigl( \frac{11}{6} -\frac{j_0}{6} + \frac{7 q_0}{6} + \frac{q_0^2}{2} \Bigr) + \nonumber\\
    &&+\, z^4 \cdot \Bigl( -\frac{25}{12} + \frac{13 j_0}{24} - \frac{23 q_0}{12} + \frac{5 j_0 q_0}{12}
    - \frac{13 q_0^2}{8} - \frac{5 q_0^3}{8} + \frac{s_0}{24} \Bigr)+\ldots\Bigr]\,.
\end{eqnarray*}

\noindent Thus, by considering the redshift definition in terms of the cosmic time ${\displaystyle \frac{d\log(1+z)}{dt}=-H(z)}$, we can rewrite $\mathcal R$ as a function of  $z$, being
\begin{equation}\label{eq: constr}
\mathcal{R} = 6 \left[ (1+z)H\,H_{z} - 2 H^2\right]\,.
\end{equation}
Hence, it is now easy to get $\mathcal{R}$ and derivatives in terms of $z$, and to evaluate the corresponding values at present time. We obtain

\begin{equation}\label{Rinz0}
\mathcal{R}_0 =\,  6H_0\left[H_{z0}-2H_0\right]\,,\quad
\mathcal{R}_{z0} =\, 6H_{z0}^2+H_0(-3H_{z0}+H_{2z0})\,,
\end{equation}

\noindent that, for present time, simply allow us to write
\begin{equation}\label{Hinz0}
\begin{split}
H_{z0}=\,& H_0(1+q_0)\,,\\
H_{2z0}=\,& H_0(j_0-q_0^2)\,,\\
H_{3z0}=\,&H_0(-3j_0-4j_0q_0+q_0^2+3q_0^3-s_0)\,,\nonumber
\end{split}
\end{equation}
where derivatives with respect to $z$ are indicated.
By assuming the presence of standard pressureless matter, ($\rho_m\propto a^{-3}$ and $P_m=0$), including baryons and cold dark matter, the modified Friedmann equation, which fixes the whole energy budget of the universe, easily reads
\begin{equation}
H^2 = \frac{1}{3} \left [ \rho_{curv} + \frac{\rho_m}{f'(\mathcal{R})} \right
]\,, \label{eq1}
\end{equation}
with the dynamical expression for $H$, i.e. $2 \dot{H} + 3H^2= - P_{curv}$. Eq. ($\ref{eq1}$) determines curvature corrections from which we infer the dark energy fluid. In other words, the fluid density, responsible for the cosmic speed up, could be rewritten as
\begin{equation}\label{eq:rhocurv}
\rho_{curv} = \frac{1}{f'(\mathcal{R})} \left \{ \frac{1}{2} \bigg[ f(\mathcal{R})  - \mathcal{R}
f'(\mathcal{R}) \bigg] - 3 H \dot{\mathcal{R}} f''(\mathcal{R}) \right \} \,,
\end{equation}
with the corresponding curvature pressure
\begin{equation}
\frac{P_{curv}}{\rho_{curv}} = w_{curv}= -\left(1 - \frac{\ddot{\mathcal{R}} f''(\mathcal{R}) + \dot{\mathcal{R}} \left [ \dot{\mathcal{R}}
f'''(\mathcal{R}) - H f''(\mathcal{R}) \right ]} {\left [ f(\mathcal{R}) - \mathcal{R} f'(\mathcal{R}) \right ]/2 - 3
H \dot{\mathcal{R}} f''(\mathcal{R})}\right)\,. \label{eq: wcurv}
\end{equation}
It is convenient, for our purposes, to work in terms of $f(z)$, i.e. the $f(\mathcal{R})$ explicitly depending on the redshift $z$. In so doing, we assume the functional dependence $\mathcal{R}=\mathcal{R}(z)$, and  find $f'(\mathcal{R}) =\,  \mathcal{R}_z^{-1}f_z$, $f''(\mathcal{R})=\,  (f_{2z}\mathcal{R}_z - f_z\mathcal{R}_{2z})\mathcal{R}_z^{-3}$. As it is discussed in  \cite{salzano1, ngo},  $f(\mathcal{R})$ models have to evade Solar System tests for General Relativity. This means that  gravitational coupling has to agree with the local observed value. This is possible if the conditions
\begin{equation}\label{f0fz0fzz0dopo}
\begin{split}
f_0=\,&2H_0^2(q_0-2)\,,\quad
f_{z0} =\,6H_0^2(j_0-q_0-2)\,,\quad
f_{2z0}=\,-6H_0^2\Big[s_0+4q_0+(2+q_0)j_0+2\Big]\,,
\end{split}
\end{equation}
hold.
From Eqs. ($\ref{f0fz0fzz0dopo}$),  numerical estimates are derived  once the CS is known from model independent fitting procedures by means of Eqs. ($\ref{gb}$).
The numerical results of our analysis are reported in Tab. I and II. They have been obtained by employing numerical outcomes for $\mathcal H_0$, determined by the use of Eq. ($\ref{dlinterminidiz}$).

\begin{table*}
\caption{{\small Table of supernovae best fits with 1$\sigma$ error bars.}}

\begin{tabular}{c|c|c|c} 
\hline\hline\hline

{\small Parameter}  &   {\small $d_L$}            &{\small $d_F$ }&{\small $d_A$ }\\ [1.5ex]
                    &   {\small $\chi^2_{min} = 0.9764$ }                     & {\small $\chi^2_{min} = 0.9812$ }& {\small $\chi^2_{min} = 0.9832$ }\\[0.8ex]
\hline\hline
{\small $q_0$}      & {\small$-0.791$}{\tiny${}_{-0.131}^{+0.151}$}      & {\small$-1.590$}{\tiny ${}_{-0.108}^{+0.109}$}& {\small$-1.645$}{\tiny ${}_{-0.112}^{+0.112}$}
                    \\[0.8ex]

{\small $j_0$}      & {\small $2.981$}{\tiny${}_{-0.312}^{+0.334}$}     & {\small $4.552$}{\tiny ${}_{-0.780}^{+0.791}$}& {\small$4.659$}{\tiny ${}_{-0.803}^{+0.757}$}
                    \\[0.8ex]

{\small $s_0$}      & {\small $10.878$}{\tiny${}_{-2.103}^{+2.295}$}  &{\small $18.526$}{\tiny ${}_{-1.960}^{+2.121}$}& {\small$19.523$}{\tiny ${}_{-3.356}^{+4.274}$}
                    \\[0.8ex]

{\small $P_{curv,0}$}     & {\small $-1.986\div-0.687$}     & {\small $-1.886\div-0.722$}& {\small$-1.984\div-0.789$}
                    \\[0.8ex]

{\small $f_{0}$}
                   & {\small $-5.844\div-5.280$} & {\small $-7.397\div-6.963$}& {\small$-7.514\div-7.066$}\\[0.8ex]

{\small $f_{z0}$}      & {\small $9.546\div11.730$}  &{\small $20.822\div28.946$}& {\small$21.678\div29.694$}
                    \\[0.8ex]

{\small $f_{2z0}$}      & {\small $-102.728\div-59.785$}  &{\small $-116.976\div-77.479$}& {\small$-133.166\div-72.456$}
                    \\[1.5ex]

\hline\hline\hline
\end{tabular}

{ The range of  measurements are inferred through the logarithmic formula, once the CS is fitted and  Eqs. \eqref{f0fz0fzz0dopo} assumed.

}

\label{table:y4}
\end{table*}

\begin{table*}
\caption{{\small Table of supernovae+BAO best fits with 1$\sigma$ error bars.}}

\begin{tabular}{c|c|c|c} 
\hline\hline\hline

{\small Parameter}  &   {\small $d_L$}            &{\small $d_F$ }&{\small $d_A$ }\\ [1.5ex]
                    &   {\small $\chi^2_{min} = 0.9704$ }                     & {\small $\chi^2_{min} = 0.9795$ }& {\small $\chi^2_{min} = 0.9796$ }\\[0.8ex]
\hline\hline
{\small $q_0$}      & {\small$-0.688$}{\tiny${}_{-0.122}^{+0.124}$}      & {\small$-1.216$}{\tiny ${}_{-0.133}^{+0.137}$}& {\small$-1.456$}{\tiny ${}_{-0.139}^{+0.142}$}
                    \\[0.8ex]

{\small $j_0$}      & {\small $2.126$}{\tiny${}_{-0.397}^{+0.412}$}     & {\small $3.486$}{\tiny ${}_{-0.811}^{+0.821}$}& {\small$3.744$}{\tiny ${}_{-0.786}^{+0.792}$}
                    \\[0.8ex]

{\small $s_0$}      & {\small $10.878$}{\tiny${}_{-2.103}^{+2.295}$}  &{\small $18.526$}{\tiny ${}_{-1.960}^{+2.121}$}& {\small$19.523$}{\tiny ${}_{-3.356}^{+4.274}$}
                    \\[0.8ex]

{\small $P_{curv,0}$}     & {\small $-1.677\div-0.725$}     & {\small $-1.6626\div-0.724$}& {\small$-1.696\div-0.758$}
                    \\[0.8ex]

{\small $f_{0}$}
                   & {\small $-5.620\div-5.128$} & {\small $-6.698\div-6.158$}& {\small$-7.190\div-6.628$}\\[0.8ex]

{\small $f_{z0}$}      & {\small $3.234\div6.612$}  &{\small $12.144\div20.316$}& {\small$15.318\div23.100$}
                    \\[0.8ex]

{\small $f_{2z0}$}      & {\small $-99.369\div-57.555$}  &{\small $-133.786\div-89.469$}& {\small$-141.916\div-77.910$}
                    \\[1.5ex]

\hline\hline\hline
\end{tabular}

{ The range of  measurements are inferred through the logarithmic formula, once the CS is fitted and  Eqs. \eqref{f0fz0fzz0dopo} assumed.

}

\label{table:y5}
\end{table*}

Our results indicate a slightly lower dark energy pressure than the one predicted by $\Lambda$CDM model, which typically is $P_{\Lambda CDM}\sim -0.7$. This fact  suggests that any curvature dark energy fluid should behave differently than a pure (effective) cosmological constant. In general,  numerical outcomes for $P_{curv}$ does not completely agree with  $\Lambda$CDM but give compatible acceleration parameters  for  $d_L$. 

However, our results show that present data are not enough to completely constrain possible violations of the duality problem. Even though it is difficult to definitively remove such a problem, our approach allows one to alleviate it by dealing with the BAO measurement as a Gaussian prior in the likelihood definition. In so doing, one significantly reduces the allowed phase space, circumscribing Gaussian errors and reducing any possible systematics. This procedure has been accurately investigated also with different typologies of priors \cite{nfn21}. In particular, through general analyses carried out by means of Monte Carlo Markov chains, it seems that more stringent intervals occur as the BAO measurement is included \cite{nfn23}. Such tighter best fit values are however not incompatible with our results which rely in viable cosmographic intervals \cite{gtm,gtm2,nfn21,nfn22}. More stringent values will be carried out through Monte Carlo approaches, also in the field of $f(\mathcal{R})$ gravity, as one can notice from first confirmations in \cite{ngo}. In addition, all cosmological distances indicate a positive, greater than 1, jerk parameter $j_0$, although they provide unexpectedly smaller acceleration parameters for $d_F$ and $d_A$ distances.

  The statistical strategy adopted to find out our cosmographic results permits one to immediately get the requested values for $q_0,j_0,s_0$. Our results are actually statistically consistent, as one argues from the corresponding chi square functions and $1\,\sigma$ errors. Those results remarkably show that $j_0$ and $s_0$ increase, as the acceleration parameter $q_0$ decreases and confirm previous studies on the cosmographic series, providing intervals of confidence in agreement with theoretical predictions. However, although all $\chi$ square parameters are nearly the same for all the fits involved, the leading results determined from $d_L$ better support theoretical predictions, since seem to be more compatible with a higher acceleration parameters, as one expects from current data \cite{gala,gtm,gtm2,nfn24}.

\begin{equation}
\label{summary}
f_{0}<0\,,\quad
f_{z0}>0\,,\quad
f_{2z0}<0\,,\quad
P_{curv}<P_{\Lambda CDM}\,.
\end{equation}
Summing up, the absolute values of each variable increases as one performs fits with $d_F$ and $d_A$. In other words, $d_F$ and $d_A$ analyses show significative departures, incompatible to the values found by $d_L$, for the CS, although they indicate compatible values of $f_{0}, f_{z0}$ and $f_{2z0}$ and give reasonable agreement with $P_{curv}$.

Keeping in mind such considerations, we are therefore able to infer the  functional form of $f(z)$, \emph{compatible} with previous results. In particular, we get
\begin{equation}\label{fxx}
f(z)\sim -\left(a+\sum_{i=0}^{\mathcal{N}}a^{-i^2}\right)+\mathcal{O}(z)\,,
\end{equation}
where  $a\equiv(1+z)^{-1}$, with $\mathcal{N}$ a truncating number, i.e. $\mathcal{N}\sim 4$. The approach fulfills the cosmographic bounds and gives  reasonable results for $f(\mathcal{R})$ reconstructions (see also \cite{ngo}). In particular,  $f(0)<0$ and  the statistically favored dependence of $f(z)$ on $z$ is an inverse series in terms of $z$.

\section{Discussion and conclusions}

In this paper, we present a method to fix constraints on \emph{curvature dark energy} in the context of $f(\mathcal{R}$) gravity by using a cosmographic analysis derived from  different cosmological distances. Cosmography is here adopted  to determine model independent bounds on observables  related to $f(\mathcal{R})$ and its derivatives. We follow this procedure by keeping in mind that each term of CS has  a precise physical meaning, and allows to infer dynamical properties of $f(\mathcal{R})$ curvature fluid. Rephrasing this point, the limits on CS allow  to recover present time bounds on the dark energy dynamics, regardless the particular cosmological model. In so doing, to alleviate the duality problem among cosmological distances, we investigate cosmography by means of the luminosity, flux and angular distances. The approach consists  in  using cosmography to fix cosmological bounds on curvature fluid and to choose a combination of causal distances in order to alleviate degeneracy among cosmological models. Determining  cosmographic results through the three mentioned distances leads to tighter intervals of measurements, showing that  values of curvature dark energy pressure rely in lower bounds than the intervals expected in a pure $\Lambda$CDM model. To this end, we used two data sets, i.e. the Union 2.1 compilation of supernovae and the baryonic acoustic oscillation measurements. In our  analysis, we   considered  the today observed value of the  Hubble parameter $H_0$ as a  prior. This choice is due to the need of alleviating the convergence problem which plagues cosmography  as soon as  cosmological data exceed the range $z>1$. Thus, we fix $H_0$ as a sort of \emph{initial condition} on the three distances, by means of supernovae only, using all the distances expanded at the lowest order of $z$, that is ${\displaystyle d_{L;F;A}\sim\frac{1}{H_0}z}$. The corresponding fitting value of $ H_0$ has been obtained in the supernovae range $z<0.4$ where  the convergence problem does not exist. This technique is developed in order to  determine better the values of $q_0,$ $j_0$ and $s_0$, without introducing any re-parametrization variable of the redshift. In addition, we get constraints on $f(z)$ derivatives up to the second order in the Taylor expansion around $z=0$ and  found that the acceleration parameter $q_0$ is almost compatible when inferred from $d_L$ while the jerk and snap parameters are only compatible with $j_0\geq1,$  $s_0>1$ respectively. These outcomes indicate again that the standard cosmological paradigm, namely the $\Lambda$CDM model, may be extended in terms of  curvature dark energy fluid. Finally, we proposed, as test function, a particular $f(z)$, able to reproduce the cosmographic bounds that  fairly well fits the Hubble diagram up to fourth order expansion in powers of $a(t)$. In future research,  we will refine this approach  by combining more cosmological tests and further distances. This approach will allow to  reduce the   phase spaces and  better constrain the free parameters of curvature dark energy.

\end{document}